\documentclass[aps,twocolumn,floatfix]{revtex4}

\usepackage{amsmath}
\usepackage{amssymb}
\usepackage{graphicx}

\begin{document}

\newcommand{\bmath}{\begin{displaymath}}
\newcommand{\emath}{\end{displaymath}}

\newcommand{\be}{\begin{equation}}
\newcommand{\ee}{\end{equation}}
\newcommand{\bea}{\begin{eqnarray}}
\newcommand{\eea}{\end{eqnarray}}
\newcommand{\non}{\nonumber\\}

\newcommand{\bsubeq}{\begin{subequations}}
\newcommand{\esubeq}{\end{subequations}}
\newcommand{\bitemize}{\begin{itemize}}
\newcommand{\eitemize}{\end{itemize}}
\newcommand{\ket}[1]{\left|{#1}\right\rangle}
\newcommand{\bra}[1]{\left\langle{#1}\right|}
\newcommand{\abs}[1]{\left|{#1}\right|}

\newcommand{\bmx}{\begin{bmatrix}}
\newcommand{\emx}{\end{bmatrix}}
\newcommand{\bsmx}{\begin{smallmatrix}}
\newcommand{\esmx}{\end{smallmatrix}}

\newcommand{\bquote}{\quotedblbase{}}
\newcommand{\equote}{\textquotedblright{ }}

\title{Dipole-dipole instability of atom clouds in a far-detuned optical dipole trap}
\author{D.\ Nagy and P.\ Domokos}

\affiliation{Research Institute of Solid State Physics and Optics, Hungarian Academy of Sciences, H-1525 Budapest P.O. Box 49, Hungary}

\begin{abstract}
The effect of the dipole-dipole interaction on the far-off-resonance optical dipole trapping scheme is calculated by a mean-field approach. The trapping laser field polarizes the atoms and the accompanying dipole-dipole energy shift deepens the attractive potential minimum in a pancake-shaped cloud.  At high density  the thermal motion cannot stabilize the gas against self-contraction and an instability occurs. We calculate the boundary of the stable and unstable equilibrium regions on a two-dimensional phase diagram of  the atom number and the ratio of the trap depth to the temperature. We discuss the limitations imposed by the dipole-dipole instability on the parameters needed to reach Bose-Einstein condensation in an optical dipole trap.

\end{abstract}

\maketitle

\section{Introduction}

The optical dipole trap provides a conservative and tightly confining trapping potential for polarizable particles. Owing to its simplicity  it has become a standard tool for manipulating neutral atoms \cite{grimm00}. Today, various dipole trap based systems serve the study of many-body problems where the atom-atom interaction is an essential ingredient. In this paper we consider far-off-resonant trapping of polarizable particles interacting  radiatively through the electromagnetic vacuum field. 

In dense samples of cold atoms, two-atom processes strongly influence the trapping. For example, collisional processes (photo-association, hyperfine ground state changing collisions, etc.) are known to result in trap losses and they can limit the maximum achievable density. Collisions usually depend heavily on the internal electronic structure of the species. For interatomic distances in the range of the optical wavelength, the atom-atom interaction is dominated by the radiative electromagnetic coupling. In this case one can distinguish two limits: the photon scattering is dominated  (i) by spontaneous emission, e.g.\ in magneto-optical traps (MOT),  and (ii)  by stimulated emission, which occurs in far-off-resonant dipole traps (FORT).  Since the MOT enabled us first to capture dense atom clouds from vapour, it was first analyzed with respect to many-body effects \cite{sesko91,townsend95}. In a MOT, the cooling laser is quasi-resonant with the atoms and the sample forms an optically thick medium. The depletion of the laser beams together with the multiple spontaneous scattering of resonant photons within the sample can  lead to  instability \cite{sesko91} and extra heating \cite{hillenbrand94,boiron96}.  At high densities, the reabsorption of photons influences the laser cooling itself \cite{smith91,ellinger94}. The combination of these effects can result in a highly nonlinear collective dynamics of the atom cloud in a MOT \cite{labeyrie06,pohl06}.  

The FORT operates at an extremely low spontaneous scattering rate, thus the effects of multiple spontaneous photon scattering are strongly suppressed. The mechanism of trapping relies on the process of absorption and stimulated emission of laser photons. This process polarizes the particles, hence the dipole force is accompanied by the dipole-dipole coupling between atoms. 

In previous works a phenomenological term proportional to the square of the atom density was introduced to describe the atom-atom interaction, e.g., the effect of collisions on the loading of a FORT \cite{kuppens00,ohara01}.  This dominates in strongly localized traps, collisional blockade can prevent us from confining even two atoms in an extremely tiny FORT \cite{schlosser02}. This paper is devoted to studying, based on a microscopic theory, the radiative interaction between atoms in a FORT. We neglect the effects of the Coulomb and exchange interaction terms that would require the use of molecular potential surfaces \cite{suominen96}.  We explore the limits of the maximum achievable density in a FORT, which arise solely from the dipole-dipole energy shift.  

Tuning the laser frequency $\omega_{\mathrm L}$ very far below the atomic resonance $\omega_{\mathrm A}$ makes it possible to eliminate the spontaneous photon scattering at a constant depth of the trap potential, since the scattering rate and the trap depth scale differently with the detuning $\Delta_A=\omega_{\mathrm L}-\omega_{\mathrm A}$. The former is proportional to $2 \gamma \Omega^2/\Delta_A^2$ while the latter to $\Omega^2/\Delta_A$, where $\Omega$ is the Rabi frequency of the atom-laser coupling, $\gamma$ is the atomic linewidth (HWHM) \cite{grimm00}.   With temperatures well below the potential depth, i.e., $\hbar \Omega^2/\Delta_A \gg k_B T$, the particles are captured in the trap for long times. The dipole-dipole energy shift scales similarly to the spontaneous scattering rate, however, it is enhanced by the atom density. Thus it may become significant as compared to the dipole trap depth when the atom number $N$   in the trapped cloud obeys $N \gamma/\Delta_A \agt 0.1 $. 

High atom densities can be an issue for many kinds of experiments with FORT's. One example is the attempt to achieve  Bose-Einstein condensation with alkali gases in optical rather than magnetic traps \cite{boiron98,barrett01}.   Another example is connected to the so-called \emph{cavity cooling} method which is suitable to complement the deep conservative potential of FORT by a damping force induced by the field of a high-finesse optical resonator \cite{domokos04,vukics05b}. Various cavity cooling setups have been realized: (i) the FORT field can be along the cavity axis and detuned from the near-resonant cavity field by several free spectral ranges \cite{mckeever03a,maunz04}; or the FORT lasers can be perpendicular to the cavity axis with photons scattered into the cavity responsible for cooling \cite{nussmann05a,nussmann05b}.  The simultaneous cooling and trapping gives rise to very long trapping times for single atoms. Cavity cooling can be applied to many atoms as well, and should work efficiently until they get localized at the antinodes of the red detuned trap field \cite{domokos02b,black03,asboth05}.

The stability of an atom cloud against dipole-dipole coupling is an issue also in  the ultracold temperature regime where the atoms form a degenerate Bose-condensate. The corresponding anisotropic potential can be included in the Gross-Pitaevki equation \cite{yi00}.  There is a stable solution for the atomic mean-field wavefunction  depending on the trap aspect ratio \cite{santos00}, or on the scattering length \cite{goral00,odell04}.  The dipole-dipole interaction gives rise to density modulations \cite{giovanazzi02}, solitons \cite{pedri05}. In these works the atoms possess a permanent dipole moment. Recently, the effects of the magnetic dipole-dipole coupling has been observed in a chromium condensate \cite{stuhler05}. The laser induced dipole-dipole coupling at low temperatures has been treated in \cite{zhang94} and was shown to lead to nonlinear atom optical effects.   In this paper, Bose-Einstein condensation is concerned inasmuch as we study whether the dipole-dipole interaction excludes the possibility of forming a condensate in an optical trap. 

The paper is organized as follows. We start by recapitulating the formulae of the dipole-dipole interaction of atoms at a microscopic level in Sec.~\ref{sec:dipolint}.  Then, in Sec.~\ref{sec:fort} we take the far-off-resonance limit and keep only the leading order terms in the small parameter $\gamma/\Delta_A$.  The basic equations of the mean-field model are introduced in Sec.~\ref{sec:MF}. These can be expressed in terms of dimensionless parameters, the scaled intensity and the scaled atom number, which define the universality of this model.  We present numerical results of the stability analysis in Sec.~\ref{sec:stability}. There we show the parameter regimes of stability and instability on phase diagrams. In Sec.~\ref{sec:bec} the relation of the dipole-dipole instability limit to the limit of quantum degeneracy (Bose-Einstein condensation) is discussed. We conclude in Sec.~\ref{sec:concl}. We put in an Appendix the derivation of  the density-enhanced terms by a systematic expansion, leading to a Hamiltonian which accounts for the two-body interaction to all order.

\section{Dipole-dipole interaction}
\label{sec:dipolint}

The theory of radiative atom-atom interaction in the presence of a driving laser field is described in several papers, e.g.\ Refs.\ \cite{ellinger94,zhang94,lehmberg70,guo95}, we will mostly use the approach presented in \cite{guo95}. 

We consider a number $N$ of atoms interacting with a single Gaussian standing-wave laser field mode along the $\hat z$ direction which has a mode function
\be
\label{eq:mode}
\vec{f}({\vec r}) = \vec{\epsilon} \cos(k_z\, z)\exp\left[-2 \pi^2\, (x^2 + y^2)/w^2\right]\; ,
\ee
where $\vec\epsilon$ is the field polarization, and the mode is paraxial, $k_z\gg2\pi/w$. The atomic transition frequency is $\omega_A$, that of the laser field mode is $\omega_L$, and the detuning is defined by $\Delta_A = \omega_L-\omega_A$. The atom-mode interaction strength is described by the Rabi frequency $\Omega$ in the position of maximum coupling. We assume an $S \leftrightarrow P$ transition with a degenerate manifold of excited states and keep the three-dimensional polarizability of the atoms. Actually the fixed field polarization selects two levels taking part in the dynamics. The atomic internal degree of freedom is described by the vectorial lowering operator $\vec{\sigma} = \sum_q \vec{\epsilon}_q{\sigma}^q$ with $q=\pm 1$ and $q=0$ corresponding to the circular and linear polarizations, respectively. The quantization axis will be defined in accordance with the choice of the field polarization $\vec{\epsilon}$.

The equation of motion for the density operator in the Markoff approximation reads
\begin{equation}
\rho = \frac{1}{i\hbar}[H,\rho] + \mathcal{L}\rho \; .
\end{equation}
In a frame rotating at the laser frequency $\omega_L$, the Hamiltonian is
\begin{multline}
\label{eq:H_total}
  H =  \sum_{n=1}^{N} \left[\frac{{\vec p}_n^{\,2}}{2m} - \hbar\Delta_A
 \vec{\sigma}_n^\dag \vec{\sigma}_n
- i\hbar\Omega\,  \vec{f}({\vec r}_n) \, \left( \vec{\sigma}_n^\dagger -
 \vec{\sigma}_n \right)\right] {}\\
 - \hbar\gamma\sum_{\substack{n,m=1 \\ n\neq m}}^N\vec{\sigma}_m^\dag
\boldsymbol{\beta}(\vec{R}_{mn})\vec{\sigma}_n \; ,
\end{multline}
where ${{\vec r}}_n$, ${{\vec p}}_n$ and $\vec{\sigma}_n$ are the position, the momentum and the polarization of the $n$th atom ($n=1,\ldots, N$). Next to the single atom terms, i.e.\ kinetic energy, internal energy, and atom-field coupling, the last term contains the induced dipole-dipole interaction energy of the atoms. Note that the natural linewidth $\gamma$ characterizes the strength of this interaction. The tensor $\boldsymbol{\beta}$ depends on the 
coordinate difference $\vec{R}_{mn} \equiv {\vec r}_m-{\vec r}_n$ of the interacting pairs of atoms. 

The dipole-dipole interaction is mediated by the broadband vacuum, and is therefore accompanied by incoherent evolution. This is represented by additional terms in  the Liouville operator  responsible for the dissipation:
\begin{multline}
\label{eq:L_total}
{\cal L}\rho =  - \gamma \sum_{n=1}^{N} \Bigl( \{\vec\sigma_n^\dagger \vec\sigma_n, \rho\} - \\
   - 2 \sum_q\int d^2{\vec u}\; N_q({\vec u})\sigma_n^q e^{-i k_A {\vec u} {\vec r}_n} \rho e^{i k_A {\vec u} {\vec r}_n} \sigma_n^{q\dagger} \Bigr) \\
   - \gamma\sum_{\substack{n,m=1 \\ n\neq m}}^N \Bigl( 
   \{\vec{\sigma}_m^\dag \boldsymbol{\alpha}(\vec{R}_{mn})\vec{\sigma}_n, \rho\} - \\
-  2 \int d^2{\vec u}\; \vec\sigma_n  {\bf N}({\vec u}) 
e^{-i k_A {\vec u} {\vec r}_n} \rho e^{i k_A {\vec u} {\vec r}_m} \vec\sigma_m^\dagger \Bigr)
\; ,
\end{multline}
where $\{\ ,\ \}$ denotes the anticommutator. The single atom terms include the spontaneous decay accompanied by momentum recoil. The tensor ${\bf N}({\vec u})=\frac{3\gamma}{8\pi}\, (1 - {\vec u}\circ{\vec u})$, and its diagonal elements   $N_q({\vec u})=\vec{\epsilon}_q^{\,\dag}  {\bf N}({\vec u}) \vec{\epsilon}_q$ are the angular momentum distribution of the spontaneous emission from the $q$-state in the excited manifold. The double sum describes the loss effect due to the dipole-dipole coupling.
 
In free space the tensors $\boldsymbol{\alpha}$ and $\boldsymbol{\beta}$ assume  the following form :
\begin{multline}
\label{eq:alpha}
\lefteqn{\boldsymbol{\alpha}(\vec{R}_{mn}) = \frac{3}{2}\Bigg\{(1 - \hat{R}_{mn}\circ\hat{R}_{mn})\frac{\sin\:kR_{mn}}{kR_{mn}}} \\
+(1 - 3\hat{R}_{mn}\circ\hat{R}_{mn})
\Bigg(\frac{\cos\:kR_{mn}}{(kR_{mn})^2} - \frac{\sin\:kR_{mn}}{(kR_{mn})^3}\Bigg)
\Bigg\},
\end{multline}
\begin{multline}
\label{eq:beta}
\lefteqn{\boldsymbol{\beta}(\vec{R}_{mn})=\frac{3}{2}\Bigg\{(1 - \hat{R}_{mn}\circ\hat{R}_{mn})\frac{\cos\:kR_{mn}}{kR_{mn}}} \\
-(1 - 3\hat{R}_{mn}\circ\hat{R}_{mn})
\Bigg(\frac{\sin\:kR_{mn}}{(kR_{mn})^2} +
  \frac{\cos\:kR_{mn}}{(kR_{mn})^3}\Bigg)
\Bigg\},
\end{multline}
where $k=|\vec{k}|\approx k_z$, $R_{mn}=|\vec{R}_{mn}|$, and $\hat{R}_{mn}$ is a unit vector along the direction of $\vec{R}_{mn}$.

The fixed field polarization selects the excited state and the atom reduces to a two-level system with \mbox{$\vec{\sigma}_n = {\sigma}_n \vec{\epsilon}$} ($n=1, \ldots, N$). The tensors 
$\boldsymbol{\alpha}$ and $\boldsymbol{\beta}$ have to be projected onto 
this particular polarization,
\be
\label{eq:proj}
\vec{\sigma}_m^\dag \boldsymbol{\beta}(\vec{R}_{mn})\vec{\sigma}_n =
\beta(\vec{R}_{mn})\sigma_m^\dag\sigma_n, 
\ee 
where $\beta(\vec{R}_{mn}) = {\vec{\epsilon}^{\, \dag}}
\boldsymbol{\beta}(\vec{R}_{mn}) {\vec{\epsilon}}$. We now evaluate this projection in two cases: for linear polarization
along $\hat x$, and for circular one in the $\hat x$--$\hat y$ plane. 

\subsection{Linear polarization}

When the polarization of the beam is linear, $\vec{\epsilon} = \hat x$, the atomic quantization axis is taken in this direction. The atomic polarization is described by the operator $\sigma_0$, and the projection given in Eq.~(\ref{eq:proj}) results in
\begin{multline}
\label{eq:beta_l_simpl}
\beta_{mn} = \frac{3}{2}\,(3\cos^2\phi_{mn} - 1)\,
\Bigg[\frac{\sin\:kR_{mn}}{(kR_{mn})^2} + 
\frac{\cos\:kR_{mn}}{(kR_{mn})^3}\Bigg] \\
 + \frac{3}{2}\, (1 - \cos^2\phi_{mn})\,\frac{\cos\:kR_{mn}}{kR_{mn}},
\end{multline}
where $\phi_{mn} = \measuredangle (\vec{R}_{mn}, \hat x)$ is the angle between 
the distance vector of the two atoms and the axis of the polarization, $\hat x$.  As this function is singular at the origin, $kR_{mn}\rightarrow 0$, later we will use rather its Fourier transform  \cite{morice95}, 
\be
\label{eq:FT_beta_lin}
\tilde\beta(\vec{k}) = \lim_{\eta\rightarrow0} \frac{3}{\sqrt{2\pi}k_L^3} \frac{(k_z^2+k_y^2)\, (k^2-k_L^2+\eta^2)}{(k^2-k_L^2)^2+2\eta^2 (k^2+k_L^2)}\; .
\ee

\subsection{Circular polarization}
If the polarization is circular, $\vec{\epsilon} = 
-\frac{1}{\sqrt{2}}(\hat{x} + i\hat{y})$, the atomic quantization axis is the field propagation direction $\hat{z}$. The atomic polarization is described by the operator $\sigma^{+1}$, and the projection Eq.~(\ref{eq:proj}) gives
\begin{multline}
\label{eq:beta_c_simpl}
\beta_{mn} = \frac{3}{4}\,(1 - 3\cos^2\theta_{mn})\,
\Bigg[\frac{\sin\:kR_{mn}}{(kR_{mn})^2} + 
\frac{\cos\:kR_{mn}}{(kR_{mn})^3}\Bigg] \\
 + \,\frac{3}{4}\, (\cos^2\theta_{mn} + 1)\,\frac{\cos\:kR_{mn}}{kR_{mn}}.
\end{multline}
The angle $\theta_{mn} = \measuredangle (\vec{R}_{mn}, \hat z)$ is the
angle between the distance vector of the two atoms and the axis $\hat z$. The Fourier transform is
\be
\label{eq:FT_beta_circ}
\tilde\beta(\vec{k}) = \lim_{\eta\rightarrow0} \frac{3}{2\sqrt{2\pi}k_L^3} \frac{(k_z^2+k^2)\, (k^2-k_L^2+\eta^2)}{(k^2-k_L^2)^2+2\eta^2 (k^2+k_L^2)}\; .
\ee

\section{Large detuning limit}
\label{sec:fort}

For red detuning ($\Delta_A<0$) the atoms are attracted to high-intensity regions of the field.  In the large detuning limit, i.e., where the magnitude of $\Delta_A$ exceeds the atomic linewidth $\gamma$ by far, $|\Delta_A| \gg \gamma$,  the laser field creates a conservative potential for the atoms. The recoil noise is so strongly suppressed that heating  plays no role on the relevant timescale of motion. It is enough to consider only the conservative part of the dynamics described by the Hamiltonian. The dissipative processes will be taken into account later by the introduction of a phenomenological temperature.

The internal electronic dynamics of the atoms consists of fast oscillations on a short timescale. This can be adiabatically eliminated to derive its effect on the external motion. The adiabatic atomic polarizations, according to Eq.~(\ref{eq:H_total}), obey the implicit equation
\be
\label{eq:impl}
\sigma_n = \frac{\Omega}{\Delta_A}f(\vec{r}_n) +
\frac{\gamma}{\Delta_A} \sum_{\substack{m=1\\ m\neq n}}^N  (\alpha_{mn} - i\beta_{mn}) \,\sigma_m \; .
\ee

To leading order in $|\gamma/\Delta_A|$, the polarization is just the first term, which gives the zeroth order of the Hamiltonian:
\begin{multline}
\label{eq:H_eff_0}
H_{\rm eff}^{(0)} =  \sum_{n=1}^{N} \left( \frac{{\vec p}_n^{\,2}}{2m} +\frac{\hbar\Omega^2}{\Delta_A} f^2({\vec r}_n)\right) \\ 
- \frac{\hbar\gamma\Omega^2}{\Delta_A^2} \sum_{\substack{n,m\\n\neq{}m}}^N
 \beta(\vec{r}_n - \vec{r}_m)f({\vec r}_n)f({\vec r}_m).
\end{multline}
This Hamiltonian $H_{\rm eff}^{(0)}$ describes the dynamics of dipolar particles, where the polarization is induced by the external driving field.  The neglected higher order  terms describe the effect of one polarized particle on the polarization of another, i.e., the {\it local field effects}. The same effect is at the heart of  the Lorenz-Lorentz refractive index of a dielectric medium. We will later discuss in which parameter regime the local field effects become significant and present a systematic derivation of an effective two-body Hamiltonian.

\section{Mean-field approximation}
\label{sec:MF}

In the following we will resort to a single-atom analysis based on a mean-field potential.
The conservative dipole trap potential is
\be
\label{eq:trappot}
V_{\rm trap}({\vec r}) = \frac{\hbar\Omega^2}{\Delta_A}f^2({\vec r}).
\ee
The mean-field dipole-dipole potential is
\be
V_{\rm dd}^{(0)}({\vec r}) = -2\hbar\gamma\frac{\Omega^2}{\Delta_A^2}f({\vec r})
\int{}d^{3}{\vec r}_2 \;p({\vec r}_2) \beta({\vec r} - {\vec r}_2)
 f({\vec r}_2),
\label{eq:mfd}
\ee
where the continuous position distribution $p(\vec r)$ was introduced. This distribution is normalized to the number of atoms in the trap $N$.  The convolution integral can be evaluated as a product in Fourier space \cite{goral00,pedri05}. 

The final step in setting up the model consists of assuming that the cloud of atoms is described by a canonical ensemble at an equilibrium temperature $T$. By this approximation we avoid describing the heating and cooling processes which lead to the steady-state. Both are slow processes as the recoil noise is strongly suppressed. The canonical distribution provides a self-consistent equation for the spatial distribution of the atoms:
\be
\label{eq:term}
p({\vec r}) = \frac{1}{\cal Z} \exp\left[-\frac{V({\vec r}, p({\vec r}))}{k_BT}\right]\; ,
\ee
where the partition function ${\cal Z}$ ensures that the integral of $p(x)$ gives the number $N$ of atoms. This  self-consistent equation for the atom distribution is the basic equation of our model, which can be solved only numerically. The same method has been used to study phase transitions of atom gases when the atom-atom interaction is mediated by a cavity field \cite{asboth05,nagy06}.

The MF model can be expressed in terms of dimensionless parameters, which amounts to the identification of the relevant quantities describing the equilibrium of a trapped cloud of atoms. The dipole trap depth (see Eq.~(\ref{eq:trappot})) is set by the intensity and the detuning, however, in the self-consistent equation (\ref{eq:term}) it is compared to the temperature. None of the above quantities appears separately, thus it is appropriate to introduce the scaled intensity 
\begin{equation}
{\cal I} = \frac{\hbar\Omega^2}{|\Delta_A|k_BT}\; .
\end{equation}
The depth of the MF dipole-dipole potential relative to the trap depth is determined by the product of the small parameter $\gamma/\Delta_A$ and the atom number $N$ through the atomic density $p(\vec r)$ in
Eq.~(\ref{eq:mfd}). The appropriate parameter is then the scaled atom number 
\begin{equation}
{\cal N} =\frac{N\gamma}{|\Delta_A|}\; .
\end{equation}
With these two dimensionless parameters, ${\cal I}$ and ${\cal N}$, the effect of all the relevant physical quantities can be described. 

\section{Stability analysis}
\label{sec:stability}

The solution of Eq.~(\ref{eq:term}) can be determined numerically by
iterating the atomic distribution as follows. Initially we take the canonical distribution of a noninteracting gas in the dipole trap.  The dipole-dipole interaction term given by the convolution integral in Eq.~(\ref{eq:mfd}) is calculated then from this distribution. It is suitable to go into Fourier-space where the convolution is a simple product. The Fourier transform of the term $\beta({\vec r})$ is known, cf.\ Eqs.\ (\ref{eq:FT_beta_lin}) and (\ref{eq:FT_beta_circ}), while the term $p({\vec r}) f({\vec r})$ is transformed by numerical Fast Fourier Transform (FFT). Due to the finite support of the distribution $\tilde p({\vec k})$ and the mode function $\tilde f({\vec k})$, the singularity of  $\beta({\vec r})$, which appears as a non-decaying Fourier transform in momentum space, is automatically regularized. On transforming the product back to real space by FFT we get the dipole-dipole term. Adding it to the dipole trap term of Eq.~(\ref{eq:trappot}) yields the 
total mean field potential which  furnishes a new atomic distribution via the canonical form in Eq.~(\ref{eq:term}). The resulting $p(x)$ can be used as the starting distribution in the next step of the iteration.  Continuing the steps of iteration until convergence, one obtains the self-consistent solution of Eq.~(\ref{eq:term}).  

The iteration does not necessarily converge: instability can be induced by large enough atom number or intensity. The iteration method suggests that the instability occurs as a collapse of the atomic cloud due to self-contraction in the center of the trap. However, in the lack of a precise modeling of the collisional processes, the collapse itself cannot be accounted for by our approach. We must limit ourselves to determining the range of convergence. 

The mean field dipole-dipole energy Eq.~(\ref{eq:mfd}) depends  on the shape of the atomic cloud. It is the pancake-shaped trap where the dipole-dipole contribution to the MF potential is negative, deepening the trap depth in the center \cite{santos00}, so that the collapse of the cloud can be expected. It vanishes in the center for a spherical atom distribution (the refinement of this statement can be found in \cite{wesenberg04}). For a cigar-shaped cloud the MF dipole-dipole energy would be positive, repelling atoms from the center, and thus the instability we discuss in the following could not occur.

\begin{figure}[ht]
\begin{center}
  \includegraphics*[angle=-0,width=7cm]{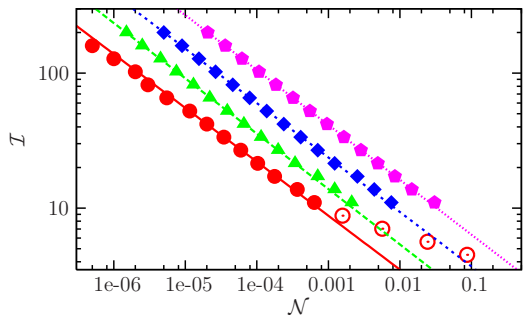}
\end{center} 
\caption{The boundary of the stability range of the atomic cloud is shown in logarithmic phase diagram of the scaled intensity ${\cal I}$ and the scaled atom number  ${\cal N}$. The field polarization is circular in the $\hat x-\hat y$ plane, and the waist is $w / \lambda$= 1.33 (circle), 2.66 (triangle),  5.33 (diamond), and 10.66 (pentagon). Straight lines represent a fit on the numerical data. As shown with empty circles, below ${\cal I}=10$ the phase boundary deviates from the corresponding fitted line ($w / \lambda= 1.33$), which indicates the effect of untrapped atoms.}
\label{fig:stability} 
\end{figure}
In Fig.~\ref{fig:stability} the stability range of the iterations is presented for circular polarization in the $\hat x$--$\hat y$ plane, and for beam waists $w=1.33$, 2.66,  5.33, and 10.66 in units of $\lambda$. On the two-dimensional plot for the scaled atom number and scaled intensity variables (``phase diagram''), border points of the stability region are shown and the convergent iterative solution of Eq.~(\ref{eq:term}) exists in the region {\it below} the points. The border points  can be well fitted by a power law dependence of the critical scaled intensity on the scaled atom number,  ${\cal I} \propto {\cal N}^{-c}$. The fit is represented by lines, the exponent  is $c = 0.40(\pm 0.01)$. Shown only for the $w/\lambda =1.33$ data with empty circles, the boundary bends away from the fitted straight line at the right-most end. This happens below a certain scaled intensity (${\cal I}<10$), when a significant portion of untrapped atoms appear.  It is not shown on the figure, however, we obtained the same results for linear polarization along $\hat x$. We also checked that the addition of an arbitrary constant to the Fourier transform $\tilde\beta({\vec k})$ does not appreciably shift the phase boundary. This justifies the neglect of any type of Dirac-$\delta (\vec r)$ potential in the Hamiltonian, e.g., the contact potential \cite{morice95} or $s$-wave scattering.

\begin{figure}[ht]
\begin{center}
\includegraphics*[angle=-0,width=7cm]{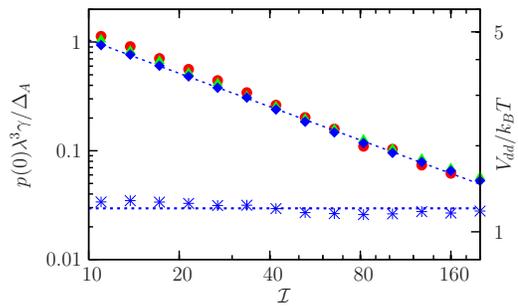}
\end{center} 
\caption{Left scale: Maximum density of the cloud, $p(0)$ in units of $1/\lambda^3  \times \Delta_A/\gamma$,  plotted against the scaled intensity ${\cal I}$ for  $w / \lambda$= 1.33 (circle), 2.66 (triangle),  5.33 (diamond). The straight line is a fit on the  $w / \lambda =5.33$ data. The scaled atom number ${\cal N}$ is set such that the system remains slightly below the critical point. Right scale: stars with a constant fit represent the MF dipole-dipole potential $V_{\rm dd} / k_BT$ calculated close to the boundary of the range of stability. The beam waist is $w/\lambda = 5.33$.}
\label{fig:rhomax}
 \end{figure}
The dipole-dipole coupling enhances the trap depth in the center and increases the atom density there. The resulting self-contraction of the cloud is counteracted by the random motion of the atoms.  We expect that instability occurs when the energy shift due to the dipole-dipole coupling exceeds the thermal energy of the atoms. The numerical approach has allowed us to confirm this expectation. In Fig.~\ref{fig:rhomax} the ratio of the dipole-dipole interaction potential and the thermal energy is plotted (with stars) on a logarithmic scale at the edge of the stable region (where still stable solution exists), and one finds  $V_{\rm dd} \approx k_B T$ closely constant. 

Accepting the instability condition $V_{\rm dd} \approx k_B T$, the critical exponent $c=0.4$ can be obtained by simple arguments. It follows from Eq.\ (\ref{eq:mfd}) that  
$V_{\rm dd}^{(0)}(0)/k_B T$ is proportional to ${\cal I}{\cal N}$ times the convolution integral. Using the dominant term of the potential for the distribution, i.e., the trap potential in harmonic approximation, the resulting  Gaussian distribution has a normalization factor proportional to ${\cal I}^{3/2}$. The remaining part is the convolution involving the function $\beta$ which is singular at the origin. Thus the main contribution must come from this small domain, the cloud size is irrelevant and the integral must be determined, at least to leading order, by the aspect ratio of the trapped cloud. Altogether $V_{\rm dd}^{(0)}(0)/k_B T \propto {\cal I}^{5/2}{\cal N}$ from which the scaling ${\cal I} \propto {\cal N}^{-0.4}$ follows.

\begin{figure}[ht]
\begin{center}
 \includegraphics*[angle=-0,width=7cm]{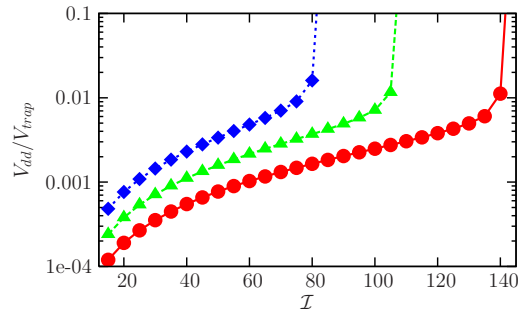}
\end{center} 
\caption{The ratio of the MF dipole potential and the total potential at the origin as a function of the scaled intensity  ${\cal I}$ for $w/\lambda = 5.33$, and at fixed values of ${\cal N} = 1$, $2$, $4 \times 10^{-5}$.}
\label{fig:diprat} \end{figure}

For consistency, let us check the validity of the effective Hamiltonian given in Eq.~(\ref{eq:H_eff_0}) which keeps only the leading order term of the dipole-dipole interaction.  Figure~\ref{fig:diprat} depicts the ratio of the MF dipole-dipole potential to the trap potential at the center of the cloud, which is the same ratio as that of the higher order terms of the polarization to the leading order one [cf.\ Eq.~(\ref{eq:impl})]. 
This ratio grows as a function of the scaled intensity until reaching the critical point which is clearly manifested in this semi-logarithmic plot. The main thing to observe is that the criticality is reached at fairly low ratio of the dipole-dipole potential to the trapping one, at a value well below $0.1$.  Therefore, in the considered parameter regime, the leading order description is enough to find the boundary of the stability domain. For other parameter regimes, where the ratio would be larger than 0.1 but still below 1, we present,  in the Appendix, a Hamiltonian which accounts for the two-body interaction to all order.

\section{Possibility of Bose-Einstein condensation}
\label{sec:bec}

The peak density is a key quantity with respect to reaching  quantum degeneracy, which happens roughly at the density $p(0) \lambda^3_{\rm deBroglie} \approx \zeta(3/2) = 2.612$ \cite{castin01}, or, expressed in another way,
\be
 \label{eq:bec}
p(0) \lambda^3  \approx 41  \left(\frac{T}{T_{\rm rec}}\right)^{3/2}\; ,
\ee
where $T_{\rm rec} = \left(\hbar^2 k^2 /m\right)/k_B$ is the recoil temperature. This condition depends, separately, on the temperature which has been, so far, embedded in the scaled intensity. Therefore, depending on the working point, i.e.\ the actual value of the  temperature and the trap depth, the order of reaching quantum degeneracy or reaching the dipole-dipole instability can be chosen. 

Obviously, both increasing the trap depth (at fixed temperature) and increasing the absolute atom number (at fixed detuning) lead to an increase of the density at the trap center. However,  the critical behavior can be induced at different peak densities depending on how it is obtained: whether by increasing ${\cal I}$ or ${\cal N}$.  We present in Fig.~\ref{fig:rhomax}  the calculated peak densities along the phase boundary, i.e., for clouds which are infinitesimally close to the boundary of the stable phase. The numerical values obtained for various waists  overlap, and can be well fitted by a power law function $1/{\cal I}$. Thus the dipole-dipole instability appears at densities 
\be
 \label{eq:dd_instab}
  p(0) \lambda^3 \approx \frac{\Delta_A}{\gamma} \frac{10}{\cal I}\; .
\ee
We recall that the dimensionless detuning $\Delta_A/\gamma$ describes the ratio of the dipole trap potential depth and the spontaneous photon scattering rate.  
Thus if the right-hand-side of the inequality (\ref{eq:dd_instab}) is smaller than that of (\ref{eq:bec}), the dipole-dipole instability prevents that the limit of quantum degeneracy could be reached. For example, with atoms at about 100 $\mu$K, 3 orders of magnitude above the (typical) recoil temperature, and for ${\cal I}=10$ (trap depth about mK), one needs a  detuning $|\Delta_A| > 10^5 \gamma$ so that the Bose-Einstein condensation point falls into the stability range of the dipole-dipole interaction. 
On increasing the scaled intensity ${\cal I}$,  the threshold density $p(0)$ for the dipole-dipole instability decreases (see Fig.\ \ref{fig:rhomax} and Eq.\ (\ref{eq:dd_instab})), which amounts to a more demanding condition on the  atom number or on the temperature needed to achieve condensation. 

\section{Conclusion}
\label{sec:concl}

We set up a simple but versatile model to study the stability of far-off-resonance trapping against the dipole-dipole interaction. The model can be adapted to various geometries, atomic species, polarizations, etc. We found that the radiative atom-atom interaction in the laser polarized gas can indeed produce an  instability of a pancake-shaped atomic cloud. When the dipole-dipole energy shift reaches the temperature, the unbalanced attraction in the trap center yields a collapse of the cloud. The condition for the instability in terms of the atom density and the temperature is obtained numerically in the form of phase diagrams. 

The density necessary  for large enough dipole-dipole shift is quite large, however, the temperature is not needed to be very low so that the instability be observable. Provided  the optical dipole trap is deep enough, the stability can be lost at moderate phase-space densities, well above the condensation threshold. On the other hand, it is also possible to find a working point, very large detuning, low temperature and not too large trap depth, where the dipole-dipole coupling itself does not prevent the gas from forming a quantum degenerate state.  The dipole-dipole interaction strongly dominates the influence of quantum statistics in near-resonant laser fields \cite{menotti99}. This is not the case for very large detunings, hence we expect that the effect of cloud instability offers parameter regimes where quantum statistical properties can be probed. This is the subject of our forthcoming research.

\begin{acknowledgments}
This work was supported by the National Scientific Fund of Hungary
(Contract Nos.~T043079, T049234, NF68736).
\end{acknowledgments}

\section{Appendix}

In the Appendix we present an effective two-body Hamiltonian $H_{\rm eff}$ that takes 
into account the possibility of close pairs of atoms in a non-perturbative manner.

Eq.~(\ref{eq:impl}) can be solved iteratively to obtain 
\begin{multline}
\label{eq:it_sol}
\frac{\Delta_A\sigma_n}{\Omega} =
f(\vec{r}_n) + \frac{\gamma}{\Delta_A} \sum_{\substack{m=1\\ m\neq n}}^N \Theta_{nm} f(\vec{r}_m) \\
+\left(\frac{\gamma}{\Delta_A}\right)^2 \sum_{\substack{m=1\\ m\neq n}}^N \sum_{\substack{l=1\\ l\neq m}}^N\Theta_{nm} \Theta_{ml} f(\vec{r}_l) \\ +
\left(\frac{\gamma}{\Delta_A}\right)^3 \sum_{\substack{m=1\\ m\neq n}}^N \sum_{\substack{l=1\\ l\neq m}}^N
\sum_{\substack{k=1\\ k\neq l}}^N\Theta_{nm} \Theta_{ml} \Theta_{lk} f(\vec{r}_k) + \ldots 
\end{multline}
where
$\Theta_{mn} = \alpha_{mn} - i\beta_{mn}$ has been introduced for the sake of 
compactness. The series includes terms of high order in the small parameter $\gamma/|\Delta_A| \ll 1$,  however, not all of them can be neglected. Such a high-order term can contain extremely large factors provided all the $\Theta_{kl}$ in it have large values, i.e., the corresponding atoms
all are close to each other (within a small fraction of the wavelength). This is because in the limit ${R}_{mn} \rightarrow 0$, the dipole-dipole coupling function gives  $\beta(\vec{R}_{mn}) \rightarrow \infty$.  However, even in this limit $\alpha$ is bounded, i.e.\ $\alpha(\vec{R}_{mn}) \rightarrow 1$, thus the real part of $\Theta$ can safely be neglected in the large detuning limit.

A systematic low-density approximation of Eq.~\eqref{eq:it_sol} can be made by considering only two-body terms.  This amounts to the assumption that any pair of atoms can be very closely spaced, however, in this case there is no third atom in their immediate vicinity. Thus if for any $n\neq m$, $\beta_{nm}\gg 1$, then for every $l=1,\ldots,N$ such that
$l\neq n$, $l\neq m$, we have $\beta_{nm}\beta_{ml} \leq 1$.  In this
approximation, we reorder the sum, and obtain two geometric series of 
$-\beta_{mn}\gamma/\Delta_A$, leading to
\begin{multline}
\label{eq:sigma_st}
\sigma_n = \frac{\Omega}{\Delta_A} \Bigg[\Bigg( 1 + \sum_{\substack{m\\ m\neq n}}
\frac{(\beta_{mn}\gamma/\Delta_A)^2}{1-(\beta_{mn}\gamma/\Delta_A)^2} \Bigg) f(\vec{r}_n) \\ 
- \sum_{\substack{m\\ m\neq n}}
\frac{(\beta_{mn}\gamma/\Delta_A)}{1-(\beta_{mn}\gamma/\Delta_A)^2} f(\vec{r}_m) \Bigg]\; .
\end{multline}
Here we have assumed that for every atom pair $n\neq m$, $|\beta_{mn}\gamma/\Delta_A|<1$. This solution, within the Markoff and the adiabatic approximations, takes into account in a non-perturbative manner the  pairs of atoms separated by small distances.

Inserting back the stationary adiabatic atomic polarization $\sigma_n$ of Eq.~(\ref{eq:sigma_st})
into the original form of the Hamiltonian of the system Eq.~(\ref{eq:H_total}) and eliminating double sums on the same ground as above, one can end up with the effective
two-body Hamiltonian:
\begin{multline}
\label{eq:H_eff_exact}
H_{\rm eff} =  \sum_{n=1}^{N} \frac{{\vec p}_n^{\,2}}{2m} + \frac{\hbar\Omega^2}{\Delta_A}\Bigg[\sum_nf^2_n\\
+ \sum_{\substack{n,m\\ m \neq n}}f^2_n
\left\{(\beta^{'}_{mn})^2 + 2 \beta^{''}_{mn} + 3(\beta^{''}_{mn})^2\right\}\\
- \frac{\gamma}{\Delta_A} \sum_{\substack{n,m\\ m \neq n}}f_mf_n
\beta_{mn}\left\{\left(1 + \beta^{''}_{mn}\right)^2 + 3 (\beta^{'}_{mn})^2
\right\}\Bigg],
\end{multline}
where $f_n = f(\vec{r}_n)$ is the mode funcion of the field at the position of the $n$th atom, and $\beta^{'}_{mn}$, $\beta^{''}_{mn}$ are defined as below:
\begin{equation}
\label{eq:beta_p1}
\beta^{'}_{mn} = \frac{\gamma}{\Delta_A}\frac{\beta_{mn}}{1 - \frac{\gamma^2}{\Delta_A^2}\beta_{mn}^2}\, , 
\end{equation}
\begin{equation}
\label{eq:beta_p2}
\beta^{''}_{mn} = \frac{\gamma^2}{\Delta_A^2}\frac{\beta_{mn}^2}{1 - \frac{\gamma^2}{\Delta_A^2}\beta_{mn}^2}\,.
\end{equation}
Note that the denominator of the expressions Eq.~(\ref{eq:beta_p1}) and Eq.~(\ref{eq:beta_p2}) should be strictly 
greater than $0$, 
due to the relation $\beta_{mn} \gamma/|\Delta_A| < 1$ that is the condition of summing up the geometric series resulting from Eq.~(\ref{eq:it_sol}).

Apart from the zeroth order effective Hamiltonian $H_{\rm eff}^{(0)}$, Eq.~(\ref{eq:H_eff_0}), 
further terms appear in Eq.~(\ref{eq:H_eff_exact}) in higher orders of $\beta_{mn}\gamma/\Delta_A$.
The new terms have two apparent meaning: the second line of Eq.~(\ref{eq:H_eff_exact}) renormalizes
the dipole trap potential (first line), while the appearing $\beta^{'}_{mn}$, $\beta^{''}_{mn}$ in the 
third line increase the strength of the dipole-dipole interaction. This increment is negligible and 
the zeroth order effective Hamiltonian, Eq.~(\ref{eq:H_eff_0}), can be used if $\beta_{mn}\gamma/\Delta_A$ is really a small parameter, however when it gets close to one, the non-perturbative form of the effective Hamiltonian,
Eq.~(\ref{eq:H_eff_exact}), should be used.


\end{document}